# Quantum vacuum and dark matter

Dragan Slavkov Hajdukovic[1]
PH Division CERN
CH-1211 Geneva 23
dragan.hajdukovic@cern.ch
[1]On leave from Cetinje, Montenegro

**Abstract**
Recently, the gravitational polarization of the quantum vacuum was proposed as alternative to the dark matter paradigm. In the present paper we consider four benchmark measurements: the universality of the central surface density of galaxy dark matter haloes, the cored dark matter haloes in dwarf spheroidal galaxies, the non-existence of dark disks in spiral galaxies and distribution of dark matter after collision of clusters of galaxies (the Bullet cluster is a famous example). Only some of these phenomena (but not all of them) can (in principle) be explained by the dark matter and the theories of modified gravity. However, we argue that the framework of the gravitational polarization of the quantum vacuum allows the understanding of the *totality* of these phenomena.

1. Introduction

Contemporary physics has two cornerstones: General Relativity and the Standard Model of Particle Physics. General Relativity is our best theory of gravitation. The Standard Model is a collection of Quantum Field Theories; according to the Standard Model, everything in the Universe is made from six quarks and six leptons (and their antiparticles) which interact through exchange of gauge bosons (photon for electromagnetic interactions, $W^\pm$ and $Z^0$ for weak interactions and eight gluons for strong interactions).

The problem is that our best physics is apparently insufficient to explain a series of major phenomena discovered in Astrophysics and Cosmology. One of the unexplained phenomena is that the gravitational field in the Universe is much stronger than it should be according to our theory of gravity and the existing amount of the baryonic matter (i.e. the matter composed from the Standard Model particles). This phenomenon is considered as a strong hint that at least one of cornerstones (General Relativity and Standard Model) must be significantly modified. Both approaches (modification of the fundamental law of gravity and the assumption that in addition to quarks and leptons there are still unknown fundamental particles named dark particles) have been studied by thousands of scientists, but a solution is still not at hand.

Recently (Hajdukovic, 2011; but see also the first appearance of the idea in Hajdukovic, 2007 and Hajdukovic, 2008)) a third way, without invoking dark matter and without invoking the modification of the fundamental law of gravity, has been proposed. In simple words, according to the Quantum Field Theory, all baryonic matter in the Universe is immersed in quantum vacuum; popularly speaking a "sea" of short living virtual particle-antiparticle pairs (like electron-positron pairs with the lifetime of about $10^{-22} s$, or neutrino-antineutrino pairs with a lifetime of about $10^{-15} s$ which is a record lifetime in the quantum vacuum). It is difficult to believe that quantum vacuum does not interact gravitationally with the baryonic matter immersed in it. In spite of it, the quantum vacuum is ignored in astrophysics and cosmology; not because we are not aware of its importance but because no one has any idea what the gravitational properties of the quantum vacuum are. In absence of any knowledge, as a starting point, we have conjectured that particles and antiparticles have the gravitational charge of opposite sign. An immediate consequence is the existence of the gravitational dipoles; a virtual pair is a gravitational dipole (in the same way as a virtual electron-positron pair is an electric dipole), that allows the gravitational polarization of the quantum vacuum. The initial study (Hajdukovic, 2011) has revealed the surprising possibility that the gravitational polarization of the quantum vacuum can produce phenomena usually attributed to dark matter. In the present paper we focus on four benchmark phenomena established by observations: (a) the universality of the central surface density of galaxy dark matter haloes (Donato et al. 2009), (b) the cored dark matter haloes in dwarf spheroidal galaxies (Walker and Penarrubia, 2011), (c) the non-existence of dark disks in spiral galaxies (Moni Bidin et al. 2010) and (d) the distribution of dark matter after collisions of clusters of galaxies (the Bullet cluster (Clove et al. 2006) being a famous example). In section 2 we give a brief review of these four phenomena and point to the known fact that only some of them (but not all of them) can in principle be explained by the dark matter and the modified theories of gravity. In section 3 we consider the same phenomena in the framework of the gravitational polarization of the quantum vacuum and argue that it is the framework in which the totality of these phenomena can be understood. Section 4 is devoted to discussion.

## 2. Four important measurements

Let us give a brief review of four observed phenomena which have become benchmark for different theories. Both, the cold dark matter model and MOND fail to explain the totality of these phenomena. The dark matter theory has more problems at small scales, while modified gravity (we take MOND as leading example) has significant problems at large scales.

*(a) Central surface density*

There is strong evidence (Donato et al. 2009) that the central surface density $\mu_{0D} \equiv r_0 \rho_0$ of galaxy dark matter haloes (where $r_0$ and $\rho_0$ are the halo core radius and central density) is nearly constant and independent of galaxy luminosity. The measured value (Donato et al. 2009) is about 140 solar masses per square parsec

$$\mu_{0D} \equiv r_0 \rho_0 = 140^{-80}_{-30} \frac{M_{Sun}}{pc^2} = 0.29 \frac{kg}{m^2} \quad (1)$$

The universality of the dark matter surface density at the core radius is a mystery for the particle dark matter but can be explained within the MOND phenomenology (Milgrom, 2009). As we will see, the gravitational polarization of the quantum vacuum obviously leads to a relation producing the numerical result (1).

*(b) Dwarf spheroidal galaxies*

Dwarf spheroidal galaxies, with a typical diameter of about 1000 light years, are the smallest galaxies observed in the Universe. For a number of reasons they are considered as an important "laboratory" for the study of dark matter distribution at the centres of galaxies. Recently, Walker and Penarrubia (2011) have accomplished the first direct measurements that reveal how densely dark matter is packed toward the centres of two nearby dwarf galaxies (Fornax and Sculptor) that orbit the Milky Way as satellites.

The measured slope

$$\Gamma \equiv \frac{\Delta \log M}{\Delta \log r} \quad (2)$$

is $\Gamma \approx 2.61$ and $\Gamma \approx 2.95$ respectively for Fornax and Sculptor galaxy. The values of $\Gamma$ in the range $2 < \Gamma < 3$, are consistent with cored dark matter halos of an approximately constant density over the central few hundred parsecs, what contradicts the cusp distribution ($\Gamma < 2$) predicted by the current cold dark matter theory. Hence, Walker and Penarrubia have provided the first direct evidence that the cold dark matter paradigm cannot account for the phenomenology of dark matter at small scales.

*(c) Dark disks*

Everyone knows that our Galaxy is immersed in a halo of dark matter (a real one if we trust the cold dark matter theory or a phantom halo according to theories of modified gravity like MOND). It is less known that in addition to the halo, our galaxy should have a dark matter disk, which is thicker than the visible galactic disk. The presence of a real dark disk is a natural expectation of the cold dark matter model (Read et al. 2008) while the presence of a phantom disk (Milgrom, 2001) is a prediction of MOND theory. The observations suggest (Moni Bidin et al. 2010) that at this point both theories are wrong; apparently, dark matter disk does not exist. As we will show in Section 3, the non-existence of dark matter disk is a natural consequence of the gravitational polarization of the quantum vacuum.

*(d) The Bullet cluster*

The observations of the Bullet cluster show the distribution of the baryonic and dark matter after collision of two clusters of galaxies.

During the collision, the galaxies within the two clusters passed by each other without interactions (because of the large distances between them), while the interacting clouds of X-ray emitting plasma have been slowed by ram pressure. Hence, two clouds of plasma are now located between the two separated clusters. The key point is that the distribution of dark matter (determined by the gravitational lensing) is centred on clusters, while the dominant part of baryonic matter is in clouds of plasma. Such a common "destiny" of dark matter and stellar components of clusters can't be explained by modified gravity where dark matter should be centred on the dominant part of the baryonic matter (i.e. on clouds of plasma). However, in the framework of the cold dark matter theory, dark matter is collisionless and it is natural that it behaves in the same way as the collisionless part of the baryonic matter.

## 3. Gravitational polarization of the quantum vacuum

*3.1 Basic ideas*

Let us assume that particles and antiparticles have the gravitational charge of the opposite sign. Consequently, a virtual particle-antiparticle pair may be considered as a gravitational dipole with the gravitational dipole moment

$$\vec{p} = m\vec{d}; \quad |\vec{p}| < \frac{\hbar}{c} \qquad (3)$$

Here, by definition, the vector $\vec{d}$ is directed from the antiparticle to the particle, and presents the distance between them. The inequality in (3) follows from the fact that the distance between virtual particle and antiparticle must be smaller than the reduced Compton wavelength $\lambdabar_m = \hbar/mc$ (for larger separations a virtual pair becomes real). Hence, $|\vec{p}|$ should be a fraction of $\hbar/c$.

If the quantum vacuum "contains" the virtual gravitational dipoles, the gravitational field of a body immersed in the quantum vacuum, should produce vacuum polarization, characterized with a gravitational polarization density $\vec{P}_g$ (i.e. the gravitational dipole moment per unit volume).

In the quantum field theory, a virtual particle-antiparticle pair (i.e. a gravitational dipole) occupies the volume $\lambda_m^3$, where $\lambda_m$ is the (non-reduced) Compton wavelength. As argued in previous papers (Hajdukovic 2010, Hajdukovic 2011) the pions (as the simplest quark-antiquark pairs) dominate the quantum vacuum and $\lambda_m$ should be identified with the Compton wavelength $\lambda_\pi$ of a pion. Hence, the number density of the virtual gravitational dipoles has a constant value

$$N_0 \propto \frac{1}{\lambda_\pi^3} \qquad (4)$$

According to equations (3) and (4), if all dipoles are aligned in the same direction, the gravitational polarization density $\vec{P}_g$ has the maximal magnitude

$$\left|\vec{P}_g\right| \equiv P_{g\,max} = \frac{A}{\lambda_\pi^3} \frac{\hbar}{c} \quad (5)$$

where $A < 1$, should be a dimensionless constant of order of unity. This may happen only in a sufficiently strong gravitational field with magnitude $g$, larger than a critical value $g_{cr}$.

The critical field $g_{cr}$ should have the same order of magnitude (Hajdukovic, 2011) as the gravitational acceleration produced by a pion at the distance of its own Compton wavelength

$$g_{cr} = B \frac{G m_\pi}{\lambda_\pi^2} = 2.1 B \times 10^{-10} \, m/s^2 \quad (6)$$

where $B$ is a dimensionless constant of order of unity. The numerical value of $g_{cr}$ is surprisingly close to the fundamental acceleration $a_0$ conjectured by MOND; in fact $g_{cr} = a_0$ implies $B \approx 0.58 \approx 1/\sqrt{3}$ and we will adopt this value for $B$ in numerical calculations. The fact that a universal critical gravitational field $g_{cr}$ appears in our theory is only a superficial similarity with MOND; in our approach there is no modification of the fundamental law of gravity for $g < g_{cr}$.

The equations (5) and (6), together with the proportionality

$$P_{g\,max} = \frac{1}{4\pi G} g_{cr} \quad (7)$$

lead to $2A = B$, i.e.

$$A \approx 0.29 \approx \frac{1}{2\sqrt{3}}; \quad B \approx 0.58 \approx \frac{1}{\sqrt{3}} \quad (8)$$

Let us note that $1/4\pi G$ plays the role of the gravitational vacuum permittivity, analogous to the vacuum permittivity $\varepsilon_0$ in electrodynamics).

As previously suggested (Hajdukovic, 2011), dark matter density may be interpreted as the density of the gravitational polarization charges.

$$\rho_{dm} = -\nabla \cdot \vec{P}_g \quad (9)$$

If we assume the spherical symmetry, (9) may be reduced to

$$\rho_{dm}(r) = \frac{1}{r^2} \frac{d}{dr}\left(r^2 P_g(r)\right) \quad (10)$$

with $P_g(r) \equiv \left|\vec{P}_g(r)\right|$.

Let us note that from the purely mathematical point of view there are three interesting possibilities: $P_g(r)$ is directly proportional to $r$, $P_g(r) = const$ and $P_g(r)$ is inversely proportional to $r$. In these particular cases, the equation (10) leads respectively to the constant volume density, constant surface density and constant radial density of dark matter, i.e.

$$P_g(r) \propto r \Rightarrow \frac{dM_{dm}}{dV} = C_1 \quad (11)$$

$$P_g(r) = const \Rightarrow \frac{dM_{dm}}{dS} = C_2 \quad (12)$$

$$P_g(r) \propto \frac{1}{r} \Rightarrow \frac{dM_{dm}}{dr} = C_3 \quad (13)$$

where $C_1$, $C_2$ and $C_3$ are some constants. Let us note that we continue to use the words dark matter, while it is not more the dark matter of unknown nature, but the effect of the rearrangement of the virtual gravitational charges in the quantum vacuum.

The mathematical possibilities (11), (12) and (13) can approximate the real physical situations. Before showing it, let us remember that in electrodynamics, the polarization density is a function of the electric field (in some cases a linear function and in some cases a non-linear function). In the case of a hypothetical gravitational polarization, the polarization density should be a function of the strength of the gravitational field.

The first important phenomenon is saturation; in a gravitational field stronger than the critical one, the magnitude of the polarization density has the constant value determined by the equation (5). This physical situation is mathematically described by (12). The examples of a region with saturation are: the central part of our galaxy, the central part of a globular cluster and a relatively large region around a star (for instance, according to (6), our Sun produces saturation in a region larger than the solar system). By the way, let us note that the saturation in the central region is not a universal property of all galaxies; for instance in the central part of a dwarf spheroidal galaxy, the gravitational field is not sufficiently strong to produce saturation.

If the gravitational field is weaker than the critical one, the polarization density should increase when the field increases and decrease when the field decreases; consequently the equation (12) cannot be used.

Outside of a distribution of the baryonic matter, the gravitational field decreases with distance; for instance it is the case outside of the saturated region in our galaxy. This case, corresponding to (13), was already studied (Hajdukovic, 2011), leading to the main result:

$$\frac{dM_{dm}(r)}{dr} = \frac{B}{\lambda_\pi}\sqrt{m_\pi M_b} \qquad (14)$$

describing a dark matter halo outside of a spherically symmetric distribution of the baryonic mass $M_b$; a result that mimics well the observed galactic dark matter halo at relatively large distances from the center of the galaxy.

Inside a distribution of the baryonic matter, the gravitational field may increase with the distance from the center; the simplest example is the gravitational field of the Earth or the Sun, which increases from the centre to the surface and decreases after that. In the particular case of baryonic distribution with a constant volume density, the gravitational field inside the distribution is directly proportional to the distance $r$ from the center. This physical situation may be approximated with (11), but later we will use a more general dependence of the form $r^x$ with $x \leq 1$ being a positive number.

In principle, every baryonic body (a planet, a star, a complex system as a galaxy, or even a single particle such as an electron) can cause gravitational polarization, i.e. the rearrangement of the virtual gravitational charges in the surrounding quantum vacuum. This is possible only in the region in which the gravitational field of the body is stronger than the gravitational field produced by other sources, what puts a natural limit to the spatial extent of the dark matter halo of a body. If the distance of a body from other bodies increases, the size of its dark matter halo should increase as well, leading to a greater quotient of dark matter and baryonic matter. The equation (14) obtained in the previous paper (Hajdukovic, 2011) supports this intuitive picture.

As an example of the baryonic distribution without spherical symmetry, let us consider a planar-like distribution (like for instance a thin galactic disk). From the mathematical point of view, the simplest case is an infinite plane with a constant baryonic surface mass density $\sigma_b$ (this is the gravitational version of an infinite plane with constant electric charge density, what is an exercise known to every student of physics). The gravitational field $\vec{g}$ produced by the plane is perpendicular to the plane, oriented towards the plane and has a constant magnitude which can be determined by a trivial application of the Gauss's flux theorem

$$|\vec{g}| = 2\pi G \sigma_b \qquad (15)$$

In a constant gravitational field $\vec{g}$ the gravitational polarization density $\vec{P}_g$ should be a constant vector and its divergence (i.e. the right-hand side of the equation (9)) is zero. Hence, while the vacuum around the considered plane is polarized, dark matter density is zero. Consequently, close to a large plane or between two large planes, there is no significant gravitational field caused by the gravitational polarization of the quantum vacuum. By the way, it leads to the conclusion that the baryonic galactic disk of our galaxy can't be accompanied by a thicker dark matter galactic disk, what contradicts the common prediction of the cold dark matter theory (Read et al. 2008) and MOND (Milgrom, 2009) . Recent studies (Bidin et al., 2010) show that there is no evidence for a dark matter disk within 4 kpc from the galactic plane, which apparently confirm our prediction.

The above considerations suggest that we may live in a Universe with a variable quotient of the baryonic and dark matter. To see it, let us imagine, that a spherical distribution of baryonic matter is somehow "deformed" to a planar-like distribution. In these two cases, a distinct observer would measure the same quantities of baryonic matter, but different quantities of dark matter!

*3.2 Gravitational field stronger than the critical value*

Let us turn back to the case of spherical symmetry. In general, there are two regions outside a distribution of the baryonic matter; the region with $g \geq g_{cr}$ and the region with $g < g_{cr}$.

The region with $g \geq g_{cr}$ is the easiest for the study; we have the estimate (5) for the maximal magnitude of the gravitational polarization density and we can use it in the equation (10), without need for a detailed understanding of the quantum vacuum, what is the major problem in the case $g < g_{cr}$. It is evident that the mathematical case (12) corresponds to the physical case when the gravitational field is sufficiently strong to produce saturation. From (5) and (10) it is easy to obtain the relation

$$r\rho_{dm}(r) = 2P_{g\,\max} = \frac{2A}{\lambda_\pi^3}\frac{\hbar}{c} \equiv \frac{A}{\pi}\frac{m_\pi}{\lambda_\pi^2} \qquad (16)$$

which explains the observed universality (1) of the central surface density and gives (using the value of $A$ determined in (8)) a numerical value in the excellent agreement with the measurements. Alternatively we may consider the measurement (1) as the experimental determination of the constant $A$ in equations (5) and (16).

Let us forget for the moment how we have obtained the result (16). Even if considered in isolation, as an ad hoc formula, it is astonishing that a universal quantity as (1) can be expressed through universal constants and mass of a quark-antiquark pair (what is roughly a pion).

According to (16) the mass of dark matter enclosed inside a sphere with radius $r$ is

$$M_{dm}(r) = Bm_\pi\left(\frac{r}{\lambda_\pi}\right)^2 \qquad (17)$$

while the acceleration produced by the dark matter has a constant value equal to the critical acceleration.

$$g_{dm}(r) = \frac{GM_{dm}(r)}{r^2} \equiv g_{cr} \qquad (18)$$

So, in the region $g > g_{cr}$, the total acceleration at distance $r$ is the sum of the acceleration $g_b(r)$ caused by the baryonic matter (and described by the Newton law), and a very small, constant acceleration (18) caused by the dark matter and oriented towards the center of the spherical symmetry. In the region $g < g_{cr}$, $g_{dm}(r)$ is not more a constant, but depends on $r$ what can be wrongly interpreted as a modification of the Newton law (a mistake included as cornerstone of the MOND phenomenology).

By the way, the additional constant sunward acceleration (18) should exist in the Solar system and affect the orbital motions of the Solar system's bodies, but in order to detect it, we must know orbits with higher accuracy (which may be not so far into the future; Page et al. 2009)

## 3.3 Gravitational field weaker than the critical value

For bodies like a star or our planet, the gravitational field becomes stronger than the critical one in less than one meter from the center of the body. Hence, the gravitational field around a star has an inner region with $g > g_{cr}$, and an outer region with $g < g_{cr}$. The region $g > g_{cr}$ should be called the region of saturation because the polarization density has a maximal magnitude. The same should be true for a Galaxy with a supermassive black hole in the center. For instance, the supermassive black hole in the centre of the Milky Way assures condition $g > g_{cr}$ at distances of more than 100 light years (without counting other baryonic matter in the central region).

The other possibility is the existence of a large central region with $g < g_{cr}$. It is possible if there is a sufficiently low baryonic mass density in the central part of a galaxy.

Let us consider a sphere filled with the baryonic matter of the volume density $\rho_b(r)$ which depends only on the distance $r$ from the centre. The gravitational acceleration produced by the baryonic matter is

$$g_b(r) = \frac{4\pi G}{r^2} \int_0^r r^2 \rho_b(r) dr \qquad (19)$$

It is obvious that an analogous relation exists for the acceleration $g_{dm}(r)$ produced by the dark matter. In the particular case of an approximately constant baryonic volume density $\rho_b(r) \equiv \rho_b$, the equation (19) leads to the direct proportionality between acceleration $g_b(r)$ and the radial distance $r$, i.e.

$$g_b(r) = \frac{4\pi G \rho_b}{3} r \qquad (20)$$

According to (20), the assumption of the direct proportionality between $P_g(r)$ and $g(r)$ means that $P_g(r)$ is also proportional to $r$, what corresponds to the mathematical case (11), describing a cored dark matter halo.

However, at this point the problem is that we do not know the properties of the quantum vacuum and in particular we do not know if for $g < g_{cr}$, the magnitude of polarization density grows with the acceleration in a linear or non-linear manner.

To be more general, let us assume a non-linear growth of the polarization density

$$P_g(r) = K r^x \qquad (21)$$

where $K$ is a constant and $x \leq 1$ a positive number. Using this form for $P_g(r)$ in the basic equation (10) and after that using the obtained result to calculate the slope (2), leads to $\Gamma = 2 + x$, i.e. $2 < \Gamma \leq 3$, as observed for dwarf spheroidal galaxies (Walker and Panarrubia, 2011)

## 3.4 The Bullet cluster

Because of the mathematical complexity, the numerical simulations are inevitable and crucial in our present day studies of dark matter. A simulation of the Bullet cluster (and some other problems) in the framework of the gravitational polarization of the quantum vacuum is an urgent task. However it is easy to see that the observed separation of dark matter and the dominant part of the baryonic matter is not a surprise.

The key question is why there is no significant presence of dark matter between the clouds of the X-ray emitting plasma. First, while the three dimensional form of clouds is not known, during the collision the clouds were not only slowed but flattened as well. And, as we have argued above, around a flattened distribution of the baryonic matter, the additional field caused by the gravitational polarization is not significant. The second important factor is that the distance between clouds is relatively small. When two baryonic masses are close enough, they compete to orient the same dipoles in different directions, what changes the gravitational polarization density and its divergence. Hence, while without the appropriate simulations a detailed picture is impossible, the absence of dark matter in the region of clouds has nothing unusual.

4. Discussion

The initial paper (Hajdukovic, 2011) has revealed an intriguingly simple rule: find the geometrical mean of the mass of a pion and the baryonic mass of a galaxy and divide it with the Compton wavelength of the pion; what you get is very close to the observed radial dark matter density in a galaxy (see the equation (14)). It was the first indication that what we call dark matter may be the result of the gravitational polarization of the quantum vacuum.

In the present paper we have revealed the additional indications; the most striking one is the result of equation (16), a universal property of galaxies (1) can be expressed through the universal constants and mass of pion what is simply astonishing. There is one point here which deserves particular attention. The Planck constant $\hbar$, so crucial in quantum theory, but absent from our theory of gravitation, appears in both equations (14) and (16) concerned with the large scale gravitational phenomena. All this suggests that the gravitational polarization of the quantum vacuum may be a serious alternative to the dark matter paradigm.

Let us clarify that our theory is not a support to MOND. Yes, there is a critical gravitational field; in a field stronger than the critical one there is saturation (i.e. the maximal gravitational polarization density), but there is no violation of the fundamental law of gravity. The fact that MOND correctly guessed the existence of a critical field is the reason for its partial success, but (in our opinion which may be wrong) the success is limited because of the misunderstanding of the physical origin of this critical field.

Let us end with one intriguing question. Are the result (16) and its consequence (17) valid at the scale of the whole Universe? The answer may be yes. Let us use in the equation (17) the radius of the observable Universe, which is estimated to be about 14 billion parsecs i.e. $\approx 4.3 \times 10^{26} m$. According to (17) the corresponding dark matter in the Universe is about $3.4 \times 10^{53} kg$ or $1.7 \times 10^{23}$ solar masses. If our estimate of the current ratio of the baryonic and dark matter in the Universe is correct, the baryonic mass of the visible universe should be $3 \times 10^{22}$ solar masses. These numbers are close to the already existing estimations (see for instance Roos, 2003). Hence, everything looks as if equation (17) is valid for the Universe as a whole. But if so, the ratio of the dark matter and the baryonic matter in the universe should grow with time (what was already pointed in the section 3.1, from a different point of view).

At the end, let us note that the gravitational properties of antimatter would be tested in CERN before the end of the current decade (Kellerbauer et al. 2008) and that the recent theoretical considerations give some support to the gravitational repulsion between matter and antimatter (Villata, 2011).

*Note added to the proof*

The same day when this paper was accepted for publication, Wilson et al (2011) have reported the observation of the conversion of virtual photons (from the quantum vacuum) into real photons. In addition to significant indirect evidence accumulated in the past decades, this is the first *direct* evidence for the existence of vacuum fluctuations. A great support to the point of view that the physics of the 21$^{st}$ Century may well be the physics of the quantum vacuum which, as revealed in our paper, may explain the phenomena usually attributed to the mysterious dark matter.